\documentclass[cameraready]{Interspeech}
 \usepackage{float}
 \usepackage{underscore}
\title{Paediatric-HGNN: A Hybrid Heterogeneous Graph Neural Network for Detecting Disfluency in Children’s Speech via Multiscale Acoustic Fusion}

\author[affiliation={1}, orcid=0000-0003-0228-6658,correspondingauthor]{Rashini}{Liyanarachchi}
\author[affiliation={3}, orcid= 0009-0003-8285-0307]{Rachael}{Mackay}
\author[affiliation={2}, orcid=0000-0002-0175-4563]{Alison}{Short}
\author[affiliation={1}, orcid=0000-0003-2200-9703]{Aditya}{Joshi}
\author[affiliation={1}, orcid=0000-0001-8015-8358]{Erik}{Meijering}
\address{
    $^1$ University of New South Wales (UNSW), Sydney, Australia \\
    $^2$ Western Sydney University, Sydney, Australia \\
    $^3$ Resourced Music Therapy, Sydney, Australia
}

\email{r.liyanarachchi\_lekamlage@unsw.edu.au}


\usepackage{tikz}
\usetikzlibrary{positioning, fit, arrows.meta, shapes.geometric, calc, shadows}
\usepackage{comment}
\usepackage{hyperref}
\usepackage{booktabs}

\begin{document}

\maketitle

\begin{abstract}
% Automated stuttering detection (ASD) systems frequently struggle with paediatric speech due to the high acoustic variability of developing voices and the nuanced distinction between core pathological stuttering and typical developmental disfluencies. This paper introduces Paediatric-HGNN, a specialized framework utilizing a Context-aware Part-whole Interaction Network (CaPIN) designed specifically for paediatric populations. Unlike traditional 1D signal processing, our architecture employs a heterogeneous graph structure that models the hierarchical relationship between lexical units (Word Nodes) and fine-grained acoustic segments (Frame Nodes). By training exclusively on a curated paediatric corpus (UCLASS and FluencyBank), Paediatric-HGNN achieves a stable weighted accuracy of 82.4\% and a Typical Disfluency F1-score of 0.386. Our results demonstrate that grounding detection in hierarchical lexical-acoustic interactions effectively captures the ``searching'' behavior of developmental revisions, providing a more robust and interpretable tool for early clinical intervention in childhood communication disorders.

% Reduced to 1000 characters for submission
Automated stuttering detection (ASD) systems struggle with paediatric speech due to high acoustic variability in developing voices and the subtle distinction between pathological stuttering and typical developmental disfluencies. We introduce Paediatric-HGNN, a framework using a Context-aware Part-whole Interaction Network (CaPIN) tailored for paediatric data. Instead of conventional 1D signal modelling, our approach builds a heterogeneous graph capturing hierarchical relationships between lexical units (word nodes) and fine-grained acoustic segments (frame nodes). Trained on curated paediatric corpora (UCLASS and FluencyBank), Paediatric-HGNN achieves 82.4\% weighted accuracy and a Typical Disfluency F1-score of 0.386. Modelling hierarchical lexical-acoustic interactions captures developmental “searching” behaviour, offering a more robust and interpretable tool for early clinical intervention.
\end{abstract}

\section{Introduction}

Stuttering is a neuro-developmental communication disorder characterised by disruptions in the forward flow of speech, which affects approximately 5\% to 8\% of children during their preschool years \cite{stutteringPercentages}. Early diagnosis is critical, yet current clinical assessments largely rely on subjective manual observations by Speech-Language Pathologists (SLPs), which can lead to high inter-rater variability and significant time costs \cite{manualSLP}. Although automated stuttering detection (ASD) models have made strides, most are optimised for adult speech through datasets like SEP-28k \cite{Sep28k}. Consequently, these models often fail to generalise to the unique acoustic and linguistic variability of children, as they have not been extensively trained or evaluated on the higher fundamental frequencies and the phonological characteristics of paediatric populations \cite{Alharbi_2018}.

% While automated stuttering detection (ASD) systems use advanced speech-language foundation models, most are optimized for adult speech (e.g., SEP-28k) and fail to generalize to the unique acoustic and linguistic variability of children.

% Paediatric speech presents a unique challenge: distinguishing between Core Stuttering (pathological blocks and prolongations) and Typical Disfluencies (developmental word repetitions and revisions). 

Paediatric speech presents a unique diagnostic challenge: unlike adult speech, early childhood language development involves a natural phase of typical disfluency with significant acoustic and linguistic overlap between core stuttering (e.g., pathological blocks and part-word repetitions) and typical disfluencies (e.g., developmental whole-word repetitions and revisions) \cite{Yairi_1999}. As differential diagnosis relies on subtle distinctions, clinicians cannot simply trust automated predictions blindly. Also, standard ``black-box" deep learning models often lack the interpretability required for clinical adoption.

Here we propose Paediatric-HGNN, a novel approach to paediatric-specific ASD using a heterogeneous graph neural network (GNN) framework named the Context-aware Part-whole Interaction Network (CaPIN). Unlike previous methods that treat speech as a 1D signal, our architecture models speech as an interaction between two distinct node types: word nodes representing the lexical intent (actual words), and frame nodes representing fine-grained acoustic segments.

Our contributions include: 1) \textbf{Paediatric-Only Specialisation:} We use a consolidated corpus of UCLASS and the Voices-CWS subset of FluencyBank, ensuring the model is trained exclusively on paediatric, spontaneous speech to address the domain shift from adult datasets. 2) \textbf{Graph-Based Lexical Integration:} By grounding detection in the ``actual word'', we address the semantic ambiguity of child speech. Crucially, this allows the model to successfully differentiate core stuttering (e.g., pathological blocks) from typical disfluencies (e.g., developmental word repetitions), a task where traditional acoustic models frequently fail. 3) \textbf{Interpretability and Validation:} We introduce a hierarchical attention mechanism that provides clinical transparency. Our approach is validated through a rigorous speaker-independent 5-fold protocol, providing a realistic benchmark for automated paediatric diagnostics.

\section{Related Work}

We prioritize paediatric-centric research (e.g., UCLASS), as adult-trained models \cite{Sep28k} often neglect child-specific physiological and developmental variability \cite{robustChild}. ASD research has transitioned from handcrafted features to deep learning: StutterNet \cite{sheike_2021} (TDNN) captures frame-level data but lacks long-range context, while ACNNs \cite{Al-banna_2022} detect multi-second disfluencies. High-dimensional frameworks like DDSS \cite{DDSS} and SSL-based Wav2Vec 2.0 adaptations \cite{Chen_2020} further improve performance, confirming that word recognition is a prerequisite for accurate detection \cite{s_Alharbi_2018}. However, state-of-the-art (SOTA) models like StuD \cite{stud_2025} remain dependent on proprietary LLM embeddings. Conversely, spatio-temporal GNNs suggest language acquisition is an interconnected network \cite{childS}. This motivates our use of heterogeneous graphs over ``black-box'' models; by distinguishing global lexical from local acoustic nodes, we enhance interpretability and capture the part-whole interactions essential for precise paediatric diagnostics.

Frontier research treats speech as a structured graph. StutterCut \cite{StutterCut_2025} uses normalised cuts for disfluency segmentation but relies on homogeneous graphs that treat all speech segments as uniform acoustic nodes. Our work bridges the gap between lexical-heavy LLM approaches and acoustic-only graph paradigms by introducing a heterogeneous graph. This architecture explicitly models the hierarchical interaction between global lexical units (word nodes) and local acoustic realisations (frame nodes), allowing the model to ground acoustic disruptions within their linguistic context.

% Trends in spatiotemporal graph neural networks (GNNs) argue for more structured representations in language acquisition, suggesting that the words a child learns are inherently connected in a network rather than isolated events \cite{childS}. This motivates the move toward heterogeneous graph structures that can explicitly model the interaction between different modalities. Unlike ``black-box'' end-to-end models, a heterogeneous approach distinguishing between global lexical nodes and local acoustic nodes offers a pathway to interpretability. Our work adopts this perspective, extending the graph-based paradigm to capture the part-whole interactions necessary for precise, clinically interpretable paediatric diagnostics.

Spatiotemporal GNNs suggest that language acquisition is a networked process \cite{childS}, motivating heterogeneous graphs. By distinguishing global lexical from local acoustic nodes, we enhance interpretability and capture the part-whole interactions essential for precise paediatric diagnostics.

\section{Methodology}

\subsection{Paediatric Speech Dataset}

% This study focuses exclusively on children's speech and utilizes data from two well-established stuttering corpora: \textit{Fluency Bank-CWS} \cite{Fluencybank} and a subset of the \textit{UCLASS} \cite{uclass} dataset.

% \textbf{FluencyBank} is a large, publicly available corpus designed for the study of fluency and disfluency phenomena. It contains recordings from speakers across different age groups and speaking tasks, including conversational interviews and reading exercises, with time-aligned fluency annotations. In this work, we selected only recordings from child speakers and combined interview and reading sessions to capture both spontaneous and controlled speech. All selected samples include validated stuttering-related annotations.

% The \textbf{UCLASS} corpus consists of speech recordings collected from speakers who stutter, spanning a wide age range and multiple speaking tasks. For this study, we used a subset of the dataset comprising recordings from \textbf{25 children}. Only samples with available and reliable annotations were included, while recordings from adult speakers or those lacking annotations were excluded.

% Overall, the final dataset is restricted to child speech and annotated samples only, ensuring consistency across both corpora. By combining data from FluencyBank and UCLASS, the dataset covers a range of speaking styles and task conditions, supporting robust evaluation of stuttering-related speech characteristics in children.

This study focuses on paediatric speech, using a consolidated corpus from the FluencyBank-CWS \cite{Fluencybank} and UCLASS \cite{uclass} datasets. To ensure a strictly child-only training distribution, we selected recordings of 25 children from UCLASS alongside the FluencyBank-CWS subset, encompassing both spontaneous conversational interviews and controlled reading tasks. Adult samples and unannotated recordings were excluded.
%Harmonizing these corpora resulted in a specialized dataset tailored to the unique acoustic, linguistic, and developmental variability of paediatric speech, supporting a robust evaluation of early-childhood stuttering characteristics.

\subsection{Annotation Schema and Label Mapping}
To ensure consistency across FluencyBank-CWS and UCLASS, we harmonised annotations into a unified 3-class taxonomy, in consultation with speech therapy experts. Focusing on clinical relevance, we grouped specific disfluency behaviours into ``Core'' versus ``Typical'' categories, as follows. \textbf{Fluent (Class 0)}: Standard fluent speech and natural pauses. \textbf{Core Stutter (Class 1)}: Pathological disfluencies including sound, syllable, and word repetitions, as well as prolongations and blocks. \textbf{Typical Disfluency (Class 2)}: Nonpathological speech disruptions such as filled pauses, phrase repetitions, and revisions.

\subsection{Feature Extraction}
% We employ a high-dimensional hybrid feature vector (Total Dim: 945) for each word node, combining deep self-supervised embeddings with physics-based and temporal features:

% \begin{itemize}
%     \item \textbf{Deep SSL Embeddings (768-dim)}: Hidden states from the last four layers of a \texttt{Wav2Vec2-base-960h} model are averaged to provide a rich, contextual linguistic representation.
%     \item \textbf{Spectrogram Texture (128-dim)}: Mel-spectrograms (64 bins) are processed via global mean and max pooling to capture the visual "signature" of stuttering events, such as the vertical stripes of repetitions or the gaps characteristic of blocks.
%     \item \textbf{Hand-Crafted Acoustic (41-dim)}: This includes 13 MFCCs, spectral centroid, bandwidth, zero-crossing rates, and onset strength to capture traditional phonetic characteristics.
%     \item \textbf{Temporal Physics (6-dim)}: Specific proxies for jitter and shimmer, including energy instability (RMS derivatives), pitch stability (via the YIN algorithm), and autocorrelation for repetition detection.
%     \item \textbf{Contextual Duration (2-dim)}: Log-scaled segment duration and a "Duration Ratio" (the ratio of current word duration to the previous word) to identify rhythmic anomalies.
% \end{itemize}

Our model extracts a 945-dimensional (945D) hybrid feature vector for each word node to capture both linguistic context and acoustic anomalies. This vector integrates \textbf{deep self-supervised contextual embeddings} from a Wav2Vec2-base-960h\footnote{https://huggingface.co/facebook/wav2vec2-base-960h} model, \textbf{mel-spectrogram textures} (via global mean and max pooling) to capture the visual signatures of disfluencies, and traditional \textbf{handcrafted acoustic features} (such as MFCCs and zero-crossing rates). Additionally, temporal physics proxy (energy instability, pitch stability via the YIN algorithm \cite{YINAlgo}, and autocorrelation) and contextual duration metrics (log-scaled segment duration and current-to-previous word duration ratios) are included to identify rhythmic and physiological disturbances.

\subsection{Hybrid Graph Construction}
\begin{figure}[!t]
    \centering
    \includegraphics[width=0.75\linewidth]{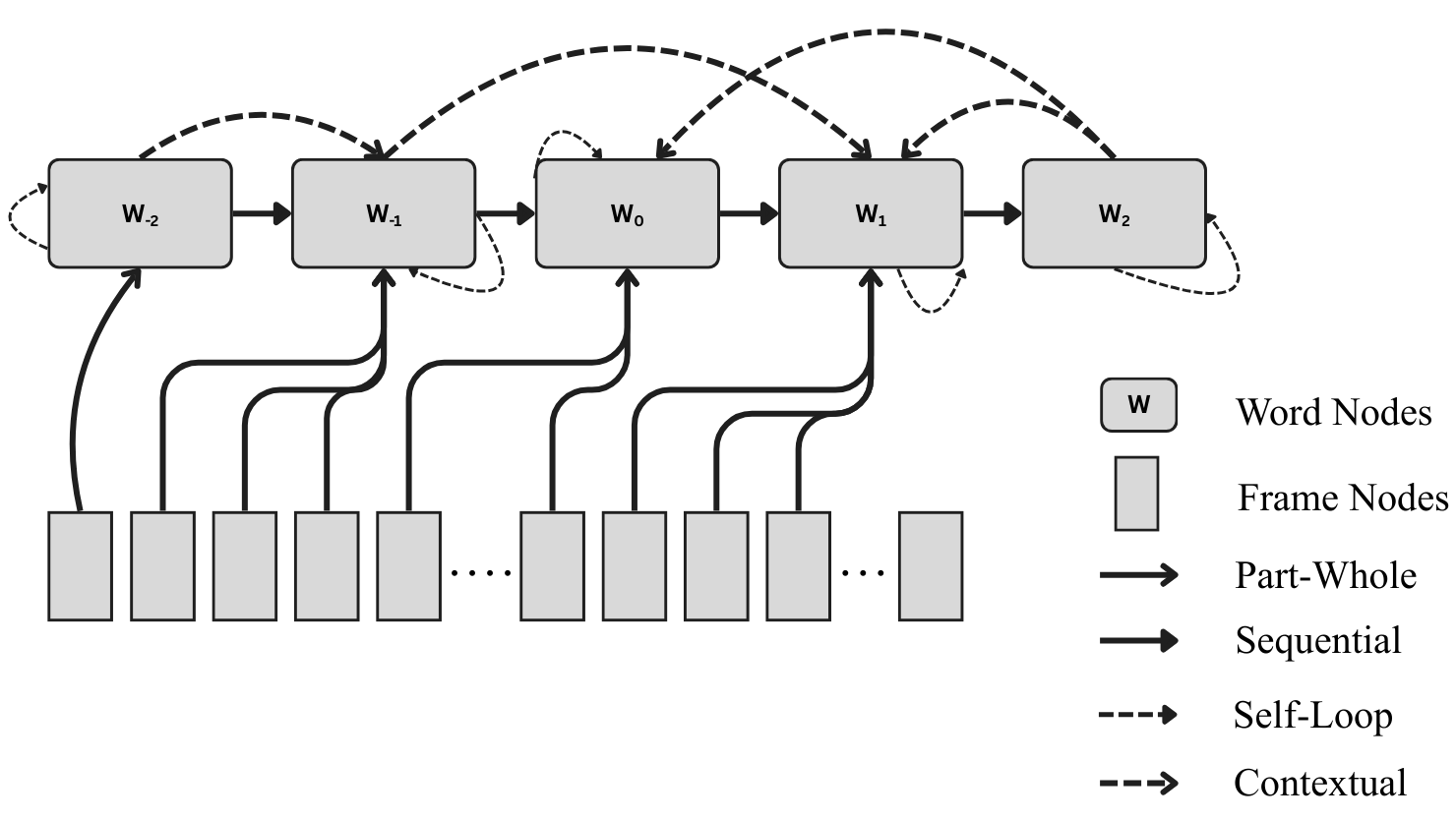}
    \caption{Heterogeneous graph architecture for hierarchical disfluency modelling. Word nodes ($W_i$) and frame nodes capture lexical intent and localised acoustic features, respectively. Graph connectivity is defined by hierarchical edges (solid open) mapping subword frames to parent words, sequential edges (solid filled) maintaining temporal flow, contextual edges (dashed open) aggregating a $\pm 2$ word neighbourhood, and self-loops (dashed filled) for feature persistence.
    %By grounding acoustic triggers within their specific lexical environment, this topology differentiates localised core stuttering from broader typical disfluencies.
    % The graph integrates fine-grained acoustic features with global lexical context by representing the speech signal through a multi-level topology where green Word Nodes ($W_n$) capture linguistic intent and blue Frame Nodes represent localized acoustic windows. Connectivity is defined by Hierarchical edges (solid black) that map sub-word acoustic disruptions to parent lexical units via temporal alignment, Sequential edges (solid purple) that maintain chronological flow, and Contextual edges (dashed orange) that bridge a $\pm 2$ word neighbourhood to aggregate local linguistic patterns. This structure, supported by Self-Loops (dashed green) for feature persistence, enables the model to differentiate between localized Core Stuttering and broader Typical Disfluencies by grounding acoustic triggers within their specific lexical environment.
    }
    \label{fig:graphStructure}
\end{figure}

The speech signal is modelled as a heterogeneous graph $\mathcal{G} = (\mathcal{V}, \mathcal{E})$ (Figure~\ref{fig:graphStructure}) to capture the hierarchical relationship between fine-grained acoustic frames and linguistic word units. \textbf{Nodes ($\mathcal{V}$)} include word nodes, initialized with the 945D hybrid vector, and frame nodes, representing acoustic windows initialized with Wav2Vec2 embeddings. \textbf{Hierarchical Edges ($\mathcal{E}_{h}$)} connect each frame node to its parent word node based on temporal alignment. \textbf{Sequential Edges ($\mathcal{E}_{s}$)} connect adjacent words to model the forward flow of speech. \textbf{Contextual Edges ($\mathcal{E}_{c}$)} provide a context window of $\pm 2$ words for nodes to aggregate information from their local linguistic neighbourhood.

% CHANGED TO BW since its mentioned to try to make it colorblind friendly

\subsection{Data Balancing and Augmentation}
To address the inherent rarity of stuttering events in spontaneous speech, we applied a targeted augmentation strategy at both the signal and graph levels. Core Stutter segments were physically transformed via segment repetition and pitch wobbling, while Typical Disfluencies were augmented with synthetic silence and energy drops to increase acoustic diversity. Following augmentation, we applied selective oversampling to further balance the distribution, replicating graphs containing Core Stutters at an $8\times$ ratio and Typical Disfluencies at a $4\times$ ratio.

\subsection{Model Architecture and Training}
\begin{figure}[!t]
    \centering
    \includegraphics[width=0.55\linewidth]{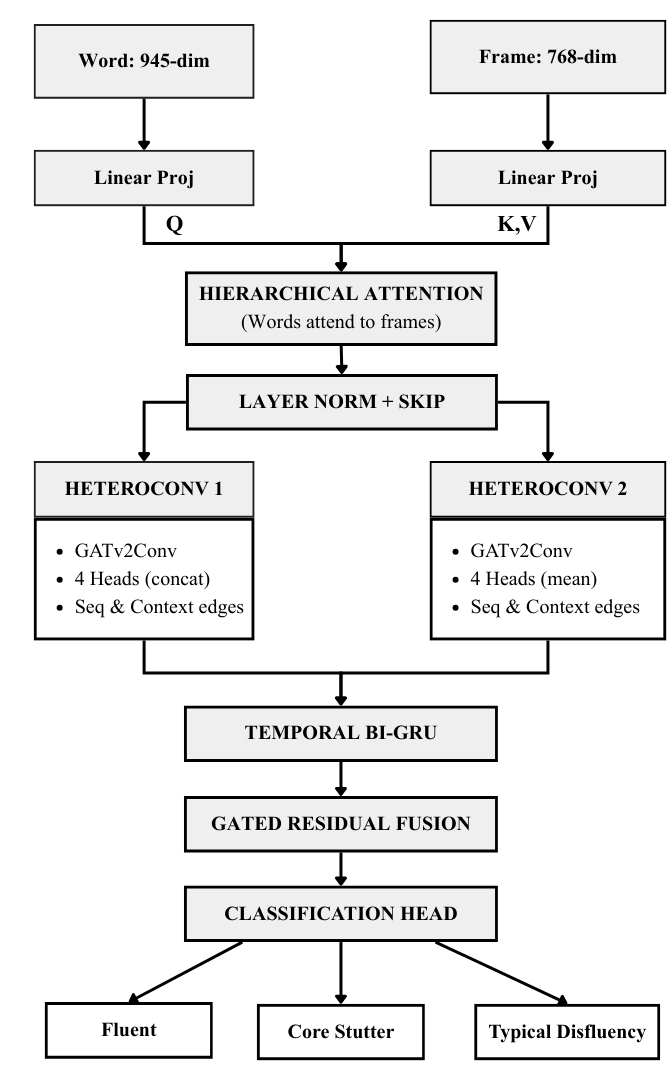}
    \caption{Paediatric-HGNN model architecture.}
    \label{fig:Paediatric-HGNN}
\end{figure}

Our proposed Paediatric-HGNN (Figure~\ref{fig:Paediatric-HGNN}) is a hierarchical heterogeneous GNN designed to capture multiscale acoustic and temporal patterns. The architecture and training pipeline consists of the following components.

\textbf{Hierarchical Graph Architecture}: The model projects word-level (945D) and frame-level Wav2Vec2 features (768D) into a shared 256D latent space. Hierarchical cross-modal attention allows word nodes (Queries) to selectively attend to relevant acoustic artifacts in their constituent frames (Keys/Values). This is followed by two stages of relational graph convolutions (GATv2Conv) across sequential and contextual ($\pm2$ words) edges to capture local and neighbourhood disfluency patterns.
    
\textbf{Temporal Integration and Gated Fusion}: To model global speech rhythm, the graph-enhanced sequence passes through a 2-layer Bidirectional GRU. A learnable gated residual fusion dynamically weights spatial GNN features against temporal RNN features: $\mathbf{H} = g \cdot \mathbf{H}_\text{RNN} + (1 - g) \cdot \mathbf{H}_\text{GNN}$. This allows prioritising local acoustic ``spikes'' for repetitions or long-range rhythmic patterns for hesitations before passing the representation to a 3-class multilayer perceptron (MLP) classifier.

\textbf{Training and Optimization}: The network is optimised using AdamW \cite{adamw} and a Focal Loss function \cite{focalLossD} ($\gamma=2$) to mitigate the impact of class imbalance. Class-specific weights $\alpha = [1, 3, 3]$ force the model to prioritise the minority (pathological) classes over the majority (fluent) class. For stable convergence, batch-level weighted sampling and dynamic thresholding post-training are used, optimising the classification boundaries to maximise F1 scores for nonfluent classes.

\subsection{Evaluation Protocol} \label{subsec:evaluationProt}
Given the data scarcity inherent in paediatric corpora, we employed a 5-fold cross-validation strategy to ensure the clinical robustness of our evaluation. To avoid information leakage, the data was partitioned such that all recordings from a specific speaker were confined to a single fold.
%This speaker-independent splitting ensures that the model is evaluated on unseen subjects, preventing the network from overfitting to speaker-specific vocal characteristics or channel artifacts.
As performance metrics we used precision, recall, F1 score, and overall accuracy, reported as the weighted averages across all 5 folds. This rigorous validation scheme provides a more realistic estimate of the model's generalisability compared to standard train-test splits often used in adult speech research.

\subsection{Experimental Baselines and Transfer Learning}
To evaluate the necessity of paediatric-specific modelling, we established a transfer learning baseline using the SEP-28k dataset \cite{Sep28k}, currently considered the SOTA benchmark for adult ASD. Our base Paediatric-HGNN architecture was pretrained on the adult SEP-28k corpus to capture generalised pathological disfluency features. This ``Adult Base'' was then fine-tuned on our harmonised paediatric dataset (UCLASS + FluencyBank-CWS). This allowed us to empirically measure the domain shift between adult motor-speech pathologies and paediatric developmental disfluencies. Beyond this transfer learning approach, we benchmarked our proposed model against several established SOTA architectures commonly applied to the UCLASS dataset, including ResNet+BiLSTM \cite{Bilstm}, StutterNet \cite{sheike_2021}, Atrous-CNN \cite{Al-banna_2022}, and Whister \cite{whister}, which represent a cross-section of 1D temporal and 2D spectral methodologies.

\subsection{Evaluation Taxonomy and Clinical Grounding}
While academic benchmarks typically demand a 4-class granularity (Fluent, Repetitions, Prolongations, Blocks), our primary evaluation uses the described 3-class clinical framework (Fluent, Core Stutter, Typical Disfluency). This is supported by the Stuttering-Like Disfluency (SLD) taxonomy \cite{Yairi_1999}, which groups part-word repetitions and dysrhythmic phonations (prolongations and blocks) into a single clinical category. In paediatric populations, the acoustic and physiological distinction between a block and a prolongation is frequently blurred due to developing speech motor control \cite{Ambrose_1999}. By consolidating these into a single Core Stutter class, we align our model with the diagnostic priorities of clinicians to distinguish persistent stuttering from typical developmental disfluencies \cite{uclass}. Nevertheless, to ensure a transparent comparison with academic benchmarks, we additionally evaluated our model using the traditional 4-class taxonomy. To ensure taxonomic alignment during comparison, the Typical Disfluency F1 scores for SOTA baselines were derived from their reported performance on ``Interjection'' or ``Other'' disfluency classes. Furthermore, our model’s performance is reported as the mean and standard deviation across the 5-fold cross-validation protocol to account for variance in the paediatric speech samples.

\begin{table}[!b]
    \centering
    \caption{Performance of Paediatric-HGNN.}
    \label{tab:results}
    \resizebox{\columnwidth}{!}{%
    \begin{tabular}{lccc}
        \toprule
        \textbf{Class} & \textbf{Precision} & \textbf{Recall} & \textbf{F1-Score} \\
        \midrule
        Fluent & $0.914 \pm 0.02$ & $0.900 \pm 0.03$ & $0.904 \pm 0.02$ \\
        Core Stutter & $0.292 \pm 0.04$ & $0.274 \pm 0.03$ & $0.280 \pm 0.06$ \\
        Typical Disfluency & $0.390 \pm 0.08$ & $0.362 \pm 0.07$ & $0.386 \pm 0.05$ \\
        \midrule
        \textbf{Weighted Average} & $0.832 \pm 0.02$ & $0.824 \pm 0.03$ & $0.826 \pm 0.02$ \\
%        \midrule
%        \textbf{Weighted Accuracy} & \textbf{82.4\%} $\pm$ \textbf{2.7\%} \\
        \bottomrule
    \end{tabular}
    }
\end{table}

\section{Results and Discussion}
\begin{table*}[!t]
\centering
\caption{Performance (F1 scores) of Paediatric-HGNN in three evaluations: 1) standard 4-class SOTA benchmark on UCLASS, 2) consolidated 3-class clinical taxonomy, and 3) ablation on the impact of adult-to-paediatric domain shift via transfer learning.}
\label{tab:comparison}
\resizebox{0.8\textwidth}{!}{%
\begin{tabular}{@{}lc@{\hspace{3em}}cc@{\hspace{3em}}c@{}}
\toprule
\multicolumn{5}{c}{\textbf{1) Standard 4-Class SOTA Benchmark on UCLASS}} \\
\midrule
\textbf{Method} & \textbf{Fluent} & \textbf{Repetition} & \textbf{Prolongation} & \textbf{Block} \\
\midrule
ResNet+BiLSTM \cite{Bilstm} & 0.52 & 0.22 & 0.28 & 0.44 \\ 
StutterNet \cite{sheike_2021} & 0.63 & 0.27 & 0.16 & \textbf{0.46} \\
Atrous-CNN \cite{Al-banna_2022} & 0.64 & 0.37 & \textbf{0.52} & \textbf{0.46} \\
Whister \cite{whister} & 0.54 & \textbf{0.47} & 0.19 & - \\
Paediatric-HGNN (Ours) & \textbf{0.90} & 0.29 & 0.39 & 0.42 \\
\midrule
\multicolumn{5}{c}{\textbf{2) Consolidated 3-Class Clinical Taxonomy}} \\
\midrule
\textbf{Method} & \textbf{Fluent} & \textbf{Core Stutter} & \textbf{Typical Disfluency} & \textbf{}\\
\midrule
ResNet+BiLSTM \cite{Bilstm} & 0.52 &  0.36 & 0.22 & \\
StutterNet \cite{sheike_2021} & 0.63 & 0.31 & 0.27 & \\  
Atrous-CNN \cite{Al-banna_2022} & 0.64 & 0.49 & 0.37 & \\
Paediatric-HGNN (Ours) & 0.90 $\pm$ 0.02 & 0.28 $\pm$ 0.06 & 0.38 $\pm$ 0.05 & \\
\midrule
\multicolumn{5}{c}{\textbf{3) Ablation on the Impact of Adult-to-Paediatric Domain Shift}} \\
\midrule
\textbf{Method} & \textbf{Fluent} & \textbf{Core Stutter} & \textbf{Typical Disfluency} & \textbf{}\\
\midrule
Pretraining SEP-28k (Adult) + Transfer Learning     & 0.88     & 0.15             & 0.08       &         \\
Paediatric-HGNN (Ours) & 0.90 $\pm$ 0.02   & 0.28 $\pm$ 0.06 & 0.38 $\pm$ 0.05 &   \\
\bottomrule
\end{tabular}
}
\end{table*}
Paediatric-HGNN, trained from scratch on paediatric data, achieved a stable overall accuracy of 82.4\% $\pm$ 2.7\%. Notably (Table \ref{tab:results}), the model demonstrated high reliability in identifying Fluent speech (F1 0.904 $\pm$ 0.02). The Typical Disfluency class reached a peak F1-score of 0.43 in specific folds (e.g., Fold 1). This confirms that the integration of lexical word nodes provides the necessary context to categorise developmental revisions that are often mislabelled in purely acoustic models.

\textbf{Failure of Adult SOTA Models for Paediatric Speech}: A key finding of this study is \textit{the significant performance degradation when utilising adult-trained models for paediatric diagnostics}. The model utilising transfer learning from SEP-28k failed to generalise to the paediatric domain (Table \ref{tab:comparison}). Despite SEP-28k being the gold standard for SOTA benchmarks, the transfer-learned model's Typical Disfluency F1 plummeted to 0.08. This catastrophic drop highlights a fundamental shift in the acoustic and linguistic domains. Adults who stutter typically exhibit stable, chronic pathological patterns. In contrast, children exhibit Typical Disfluencies (phrase repetitions and revisions) that are cognitive-linguistic in nature. Adult-trained models frequently misclassify these developmental learning phases as either fluent speech or pathological blocks, leading to poor clinical utility for early intervention.

\textbf{Analysis of Paediatric Disfluency Detection}:  A primary objective of this research was to address the acoustic ambiguity between pathological stuttering and developmental disruptions. The model achieved a mean Typical Disfluency F1 of 0.386 $\pm$ 0.05. This performance highlights the CaPIN framework's efficacy in utilising lexical context to differentiate non-pathological revisions from ``Core'' stuttering. Regarding pathological speech, the model yielded a Mean Core F1 of 0.280 $\pm$ 0.06. Although identifying Core Stutters (blocks and prolongations) in paediatric speech remains a challenge due to the lower prevalence of these events in spontaneous corpora, the model showed steady improvement during training, with individual folds peaking at a Core F1 of 0.318.

% \subsection{Interpretability and Clinical Utility}
% The "white-box" nature of the heterogeneous graph allows for granular inspection of classification triggers. By analyzing the hierarchical attention between Word and Frame nodes, we observed that: \textbf{(1) Core Stutters }typically activated Frame nodes associated with abrupt energy variations and specific spectral signatures.\textbf{ (2) Typical Disfluencies} were identified through broader temporal Word node durations and the linguistic context provided by neighboring nodes in the graph.

\begin{figure}[!t]
    \centering
    \includegraphics[width=0.75\linewidth]{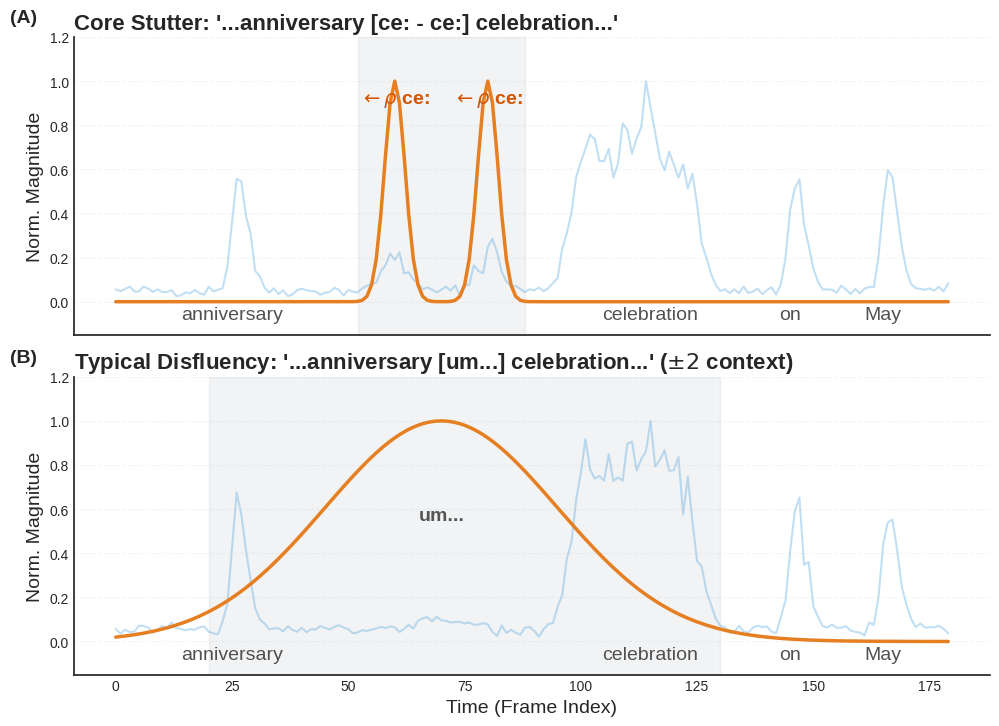}
    \caption{Interpretability analysis of hierarchical attention weights ($\phi$).
    (A) \textbf{Core Stutter}: The model exhibits sharp, localised attention on micro-level energy spikes corresponding to lengthened repeated segments ($\leftarrow \rho$ ce:), effectively ignoring the surrounding fluent context. 
    (B) \textbf{Typical Disfluency}: In contrast, the model distributes attention across a $\pm 2$ word context (e.g., ``anniversary'' to ``celebration'') to identify the linguistic planning phase associated with the filler ``um...'', demonstrating context-aware discrimination.}
    \label{fig:Inter}
\end{figure}

\textbf{Interpretability and Clinical Utility}:
The ``white-box'' nature of our CaPIN framework allows clinicians to inspect the hierarchical attention weights. For Core Stutters, the model attends heavily to micro-level energy spikes in the frame nodes (Figure \ref{fig:Inter}A). For Typical Disfluencies, attention is distributed across the word nodes and their $\pm 2$ word context, reflecting a ``searching'' behaviour rather than a motor-speech blockage (Figure \ref{fig:Inter}B). This transparency is essential for adoption by SLPs, as it provides a rationale for automated diagnosis.

\textbf{Ablation Study}: We compared the full Paediatric-HGNN against three variants: 
1) \textbf{w/o Attention}: Replaces frame-to-word attention with naive mean-pooling; 
2) \textbf{w/o Context}: Removes the $\pm 2$ word neighbourhood connections ($\mathcal{E}_{c}$); 
3) \textbf{w/o Gated Fusion}: Uses standard addition for spatial and temporal features instead of the learnable gate $g$. The 5-fold cross-validation results (Table \ref{tab:ablation}) show that the \textbf{Typical Disfluency} class is most sensitive to removal of contextual edges, with F1 dropping from 0.386 to 0.287. This confirms that developmental revisions require surrounding linguistic context for accurate identification. Conversely, the \textbf{Core Stutter} F1 falls significantly (to 0.213) without the frame-to-word attention mechanism, proving that hierarchical attention is vital for localising micro-level acoustic anomalies such as blocks. The gated fusion ensures the model adaptively weights these spatial and temporal features, providing the highest overall stability.

\begin{table}[!t]
\centering
\caption{Results of the ablation study (5-fold averages).}
\label{tab:ablation}
% \resizebox{\columnwidth}{!}{%
\resizebox{0.9\columnwidth}{!}{%
\begin{tabular}{@{}lccc@{}}
\toprule
\textbf{Variant} & \textbf{Core F1} & \textbf{Typical F1} & \textbf{Accuracy} \\ \midrule
\textbf{Full Paediatric-HGNN} & \textbf{0.280} & \textbf{0.386} & \textbf{82.6\%} \\ \midrule
w/o Attention            & 0.213         & 0.361          & 80.4\%          \\
w/o Context              & 0.263         & 0.287          & 80.7\%          \\
w/o Gated Fusion         & 0.248         & 0.337          & 81.3\%          \\
\bottomrule
\end{tabular}}
\end{table}

\section{Conclusion}
Paediatric-HGNN is a novel, heterogeneous graph-based framework specifically engineered for paediatric ASD. Our findings empirically demonstrate that, while SOTA models trained on adult corpora (e.g., SEP-28k) achieve high performance in chronic disfluency tasks, they are fundamentally ill-suited for the unique acoustic and linguistic variability of paediatric speech. This was evidenced by our transfer learning experiments, where Typical Disfluency detection collapsed to an F1 score of 0.08 when utilising an adult-pretrained backbone. In contrast, our paediatric-specialised architecture achieved a stable weighted overall accuracy of \textbf{82.4\%} and a Typical Disfluency F1 score of \textbf{0.39}. By modelling speech as a hierarchical interaction between lexical intent and acoustic realisation, Paediatric-HGNN successfully decouples pathological stutters from developmental revisions. This research underscores the critical necessity of specialised paediatric corpora and context-aware architectures to support objective, early-intervention diagnostics in speech-language pathology. The limitations of adult SOTA models highlight the need for focused research into automatic stuttering detection for child speech.

% \section{Conclusion}
% This paper introduced Paediatric-HGNN, a novel approach to paediatric-specific stuttering detection using a Context-aware Part-whole Interaction Network. By modeling speech as an interaction between lexical intent and acoustic realization, the model achieved a stable weighted accuracy of $82.4\%$. Our results confirm that grounding acoustic detection in "actual word" nodes improves the categorization of developmental revisions, as evidenced by a mean Typical Disfluency F1-score of $0.386$.

\section{Generative AI Use Disclosure}
The authors used Generative AI to edit and polish the manuscript to improve grammatical accuracy and readability. No significant part of the technical content, experimental design, or data analysis was produced by generative AI tools. All authors have reviewed the final manuscript and remain fully responsible for its contents.

\bibliographystyle{IEEEtran}
\bibliography{mybib}

% Requires: \usepackage{booktabs}

\end{document}